\documentclass[authoryear,preprint,review,12pt]{elsarticle}

\usepackage{amssymb}

\usepackage{color,soul} 

\usepackage{lineno}

\journal{Icarus}

\begin{document}


\begin{frontmatter}



\title{Micro-meteoroid seismic uplift and regolith concentration on kilometric scale asteroids}



\author[ISAE,T3]{Raphael F. Garcia}
\author[ISAE]{Naomi Murdoch}
\author[ISAE]{David Mimoun}

\address[ISAE]{Institut Sup\'{e}rieur de l'A\'{e}ronautique et de l'Espace; 10, avenue Edouard Belin, 31055 Toulouse, France}
\address[T3]{Universit\'e de Toulouse; UPS-OMP; IRAP; Toulouse, France}


%
%

\begin{abstract}
Seismic shaking is an attractive mechanism to explain the destabilisation of regolith slopes and the regolith migration found on the surfaces of asteroids \citep{Richardson2004,Miyamoto07}.  Here, we use a continuum mechanics method to simulate the seismic wave propagation in an asteroid. Assuming that asteroids can be described by a cohesive core surrounded by a thin non-cohesive regolith layer, our numerical simulations of vibrations induced by micro-meteoroids suggest that the surface peak ground accelerations induced by micro-meteoroid impacts may have been previously under-estimated.
Our lower bound estimate of vertical accelerations induced by seismic waves is about 50 times larger than previous estimates.
It suggests that impact events triggering seismic activity are more frequent than previously assumed for asteroids in the kilometric and sub-kilometric size range.
The regolith lofting is also estimated by a first order ballistic approximation.
Vertical displacements are small, but lofting times are long compared to the duration of the seismic signals.
The regolith movement has a non-linear dependence on the distance to the impact source which is induced by the type of seismic wave generating the first movement.
The implications of regolith concentration in lows of surface acceleration potential are also discussed.
We suggest that the resulting surface thermal inertia variations of small fast rotators may induce an increased sensitivity of these objects to the Yarkovsky effect. 
\end{abstract}

\begin{keyword}


Asteroids, surfaces \sep Geophysics \sep Impact processes \sep Seismic waves \sep Asteroids, dynamics
\end{keyword}

\end{frontmatter}



%
%

\section{Introduction}

In-situ observations of asteroids have highlighted the complexities and variations in the small body surface environments \citep{Cheng1997,Kawaguchi2003}.  In addition to finding each of these bodies to be regolith-covered, there is strong evidence that this regolith is very complex and active. Evidence of regolith motion has been observed on several asteroids, including landslides \citep{thomas02, veverka01a, robinson02, sierks11}, crater erasure \citep{veverka01, saito06, fujiwara06}, dust ``ponds'' \citep{veverka01, robinson01, Cheng02} and particle size segregation \citep{yano06}. 

As impact-induced seismic shaking may trigger global-scale granular processes in the dry, vacuum, microgravity environment \citep{Miyamoto2007}, it is, therefore, an attractive mechanism to explain the destabilisation of regolith slopes and the regolith migration found on the surfaces of asteroids. 
The idea of global seismic activity resulting from large impacts and the consequences on regolith motion is not a new concept. For example, \cite{Asphaug93} use hydrocodes to study the seismic resurfacing of regolith at the site of large hyper-velocity impacts. \cite{greenberg94,greenberg96} make reference to the idea of global jolting resulting from large impacts and suggest that this explains asteroid morphologies and crater populations. In addition to studying high-energy impacts, 
\cite{Housen81} consider the asteroid surface evolution produced by lower energy, non-catastrophic impact events. 
However, the focus is on the local consequences of the cratering process (e.g., burial by ejecta or excavation) rather that the global consequences of a small impact.

Here, rather than looking at the global consequences of high-energy impacts or the behaviour of regolith at the site of the impacts, we consider the motion of the non-cohesive regolith layer across the entire surface as a result of small micro-meteroid impacts. Several models have already estimated the seismic activity resulting from meteoroid impacts \citep[e.g.][]{Cheng02, Richardson2004}. In this study, we first present a new method for simulating the seismic wave propagation in an asteroid and then we consider the consequences of micro-meteoroid impacts on the regolith mobility.

In the following, we will assume that small asteroids are composed of a cohesive, or monolithic, core surrounded by a layer of non cohesive, or loose, regolith material.
We do not speculate on the type of cohesive forces at work in the cohesive core, or if this core is in fact composed of piles of blocks of various sizes, but simply assume that the seismic waves considered in this study do not induce irreversible deformations of the core.
In addition, only small amplitude seismic waves created by micro-meteoroid impacts are studied, in order to avoid irreversible deformations of the core.
With such assumptions, the propagation of seismic waves in the core can be described by continuum mechanics.
This assumption does not prevent diffraction of seismic waves by heterogeneities, voids or fractures.
\newline
The surrounding non cohesive layer is defined as a layer of loose material experiencing plastic (irreversible) deformations under the action of the seismic waves considered here. 
In particular, this layer is assumed to be able to detach from the cohesive core and subsequently experience lofting when the vertical accelerations induced by seismic waves exceed the local vertical surface acceleration.
The migration of loose regolith materials has already been observed on asteroids such as Itokawa, Eros and Lutetia \citep[e.g.][]{Miyamoto2007,Robinson2002,Sierks2011}
\newline 
This study presents continuum mechanics seismic wave propagation simulations, and suggests that the surface vertical accelerations induced by micro-meteoroid impacts may have been under-estimated.
As a consequence, we argue that impact-induced regolith motion may be more common than previously thought.
We quantify the regolith lofting by presenting first order estimates of vertical displacements and lofting time based on a ballistic approximation. 
Finally, we consider the implications of regolith concentrations in surface acceleration potential lows.

\section{Seismic modelling}

\subsection{Objective and previous studies}
The objective of our modelling is to provide an estimate of peak ground accelerations and velocities induced by direct waves for small impacts and various asteroid models in the sub-kilometric size range.
The challenge in this estimate is the proper modelling of high frequency signals, because for small impactors (mass $<$ 100 g) and small asteroids (diameter $<$ 1 km) both the excitation at the source and the elastic response of the medium present dominant frequencies larger than 20 Hz.
Examples of power spectral densities of acceleration impulse responses for different asteroid sizes and different seismic wave attenuations are presented in Figure \ref{PSD}.(a).
For asteroid sizes smaller than 1 km diameter, even at low quality factors (high attenuation) the impulse response is dominated by frequencies larger than 20 Hz.
This is due to the fact that propagation distances are small.
\newline
Most of the computations of seismic shaking effects on asteroids have been performed up to now through empirical or analytical relations scaled to large amplitude terrestrial explosions \citep{Cheng02,asphaug08}, or through hydrocode simulations \citep{Richardson2004,Walker2004}.
The majority of the studies were focused on the irreversible ground deformations due to large impactors \citep{Asphaug93,greenberg94,greenberg96}.
\newline
Two teams performed hydrocodes simulations of impacts on asteroid diameters as small as 1 km \citep{Richardson2005,Walker2004,Walker2006}.
\cite{Richardson2005} suggests that small impactors may generate peak ground accelerations larger than local gravity on small asteroids.
However, their hydrocode simulations are limited by their high frequency resolution limit at least above 50 Hz, and comparison of synthetic displacement records with theroretical Green functions (see Fig. 8 of \cite{Richardson2005}) suggest that the computed asteroid responses are low passed versions of theoretical one with a corner frequency around 20 Hz.
\cite{Walker2006} performed hydrocode simulations up to 800 Hz, with a proper quantification of an explosive source \citep{Walker2004}.
However, the attenuation model and the amount of high frequencies removed by the hydrocodes limitations are not fully described.
Moreover, the study was focused on the Eros case which falls outside our size range.
\newline
Previous simulations of seismic wave propagation in a cohesive object with continuum mechanics seismic wave equations \citep{Martin2008,Blitz2009b} predict, for objects of diameter inferior to 20 km, vertical accelerations at least two order of magnitudes larger than the local gravity for micro-meteoroid impacts (10 grams at 6 km/s).
These simulations were performed by a spectral element numerical method including wave scattering by a network of fractures.
Outputs of these models show that, despite the strong scattering of the wave field, the peak vertical ground acceleration is due to the coherent direct waves.
However, the model complexity and large computation times make it difficult to explore various asteroid sizes and shapes with such simulations.
That is why we present here a proper modelling of the high frequency elastic source signal and asteroid response through seismic normal modes computation.

\subsection{modelling tools and hypothesis}
Here, simulations of surface accelerations induced by micro-meteoroid impacts are performed by normal mode computations \citep{Zhao93,Zhao95} on spherical bodies of various radii. These computations do not reproduce the scattered wave field, but do allow an estimation of the maximum peak ground acceleration due to the direct waves.
Because the propagation distances are small, and assuming that the material remains cohesive, the scattered wave field will generate peak ground accelerations smaller than the direct waves \citep{Larose05,Martin2008,Blitz2009b}.
So, our estimates of peak ground accelerations will be valid if the energy loss of direct waves due to scattering is properly taken into account by our modelling of seismic attenuation.
\newline
A simple homogeneous internal structure model is tested with characteristics corresponding to Moon mega-regolith: $\rho=1500~kg/m^3$, $V_{P}=900~m/s$, $V_{S}=400~m/s$ \citep{Cooper1974,Horvath1980,Beyneix06}. This model can be viewed as a worst case in terms of accelerations because changing from spherical to elliptical shape would focus the seismic energy in some particular places, and adding a low velocity layer at the surface would trap the seismic waves in this layer, increasing the surface accelerations.
\newline
The source is described by an impact of a 10g projectile at 6 km/s at vertical incidence.
Previous experimental and numerical studies of such small hypervelocity impacts favour an equivalent seismic moment tensor close to the one of an explosion \citep{Walker2004}.
The total seismic moment for such an impact is estimated to be 1.8e+6 N.m from numerical experiments by \cite{Walker2004} in the energy range considered here.
This value is about 10 times smaller than the estimate from \cite{Edwards2008}, and 10 times larger than the estimate from \cite{Collins2005}.
These large variations are due to two poorly known parameters: the seismic amplification factor and the seismic efficiency.
The seismic amplification factor (ratio of the incident impulse to the impulse transfered to the target) can vary from 0.5 to 20 depending mainly on the properties of the target \citep{McGarr69}.
The seismic efficiency (part of the source energy converted into seismic waves) varies from 10$^{-6}$ to 10$^{-2}$ depending of the impactor and target properties \citep{Collins2005,Kedar2012}, and impact energy \citep{Shishkin07}.
In our simulations, the seismic moment is set to the lowest estimated value of 3.77e+5 N.m \citep{Collins2005}.
\newline
Another important parameter is the equivalent seismic source duration of the impact, because it quantifies the time range in which the energy is concentrated.
The shorter the source duration, the larger the peak ground accelerations.
Following scaling relations presented by \cite{Lognonne2009}, and using a conservative seismic efficiency of 10$^{-5}$, we predict a short source duration of 8 ms, concentrating the energy in a short time period. 
\newline
Finally, the frequency content of the signal is strongly influenced by the seismic attenuation.
As the interiors of asteroids are cold, the instrinsic wave attenuation is expected to be very low, and the attenuation of direct waves mainly due to losses by scattering, by analogy with lunar observations.
In addition, recent impact experiments suggest a low quality factor at high frequencies in the loose regolith layer at the source \citep{Kedar2012}. In order to be conservative in our estimates we tested low quality factors of 100 and 20 to mimic the scattering attenuation of the direct waves.

\subsection{modelling results}

Figure \ref{PSD}.(b) presents the envelopes of surface accelerations at 90$^{\circ}$ epicentral distance to a micro-meteoroid impact (half way point between the impact and the antipode). At low quality factors, the surface accelerations remain during a long time and pass above the maximum surface gravity during tens of seconds. Decreasing the quality factor to values as low as 20 reduces strongly the duration of the shaking. However, the energy lost in scattering will arrive later, as observed for lunar impacts \citep{Latham1970,Duennebier1974}.
In any case, even with a high attenuation, the first direct wave still generates peak ground accelerations above the gravitational acceleration.

In order to explore various sizes of asteroids, the computations were repeated for asteroid diameters from 100m to 1km.
The peak ground accelerations and velocities are presented in figure \ref{simu_sismo2}.
Our computations up to 90$^\circ$ epicentral distance for all asteroid models overlap.
It demonstrates that up to this distance the antipodal refocusing of the waves can be neglected, and it cross validates our normal modes computations for the various asteroid models.
Except for the largest asteroids and lowest quality factor, the peak ground accelerations exceed {\bf the maximum} surface gravity over the largest part of the asteroid surface. However, the peak ground velocities do not exceed the escape velocity, demonstrating that these seismic vibrations generate regolith movements, but do not allow ejection of material.
As observed on figure \ref{simu_sismo2}, peak ground accelerations and velocities vary more with distance to the source and with quality factor than with asteroid size.
However, even for such simple internal structure models, it is not possible to provide a simple scaling law as a function of distance and quality factor.

\subsection{Comparison with previous estimates}
Figure \ref{simu_sismo2}.(a) presents comparisons of estimated peak ground accelerations between our results and previous ones by \cite{Cheng02} and \cite{Richardson2005}.
The simple formula presented by \cite{Cheng02} was scaled on large amplitude terrestrial explosions, and is valid only in the near field range.
However, it validates the fact that our peak accelerations are lower bound estimates.
The estimates referring to \cite{Richardson2005} are based on an average of seismic energy over the whole asteroid, that's why they do not depend on distance, but depend on asteroid size.
Their computations assume no attenuation, a seismic efficiency of 10$^{-5}$ and a dominant frequency of 30 Hz (equations (2) and (5) of \cite{Richardson2005}).
Similar computations were performed with the seismic diffusion model (equation (14) of the same study), but due to small asteroid sizes compared to the minimum diffusion length estimate (125 m), the results are almost identical.
Due to the averaging over the asteroid volume the peak acceleration values are lower than our estimates, in particular at short distances.
Consequently we find that the minimum impactor size necessary to trigger regolith movement is even smaller than the estimate of \cite{Richardson2005}.

The micro-meteoroid events investigated here have a mean period between impacts smaller than 1 year/km$^2$ for impactors more massive than 10 grams \citep{Richardson2005}. 
Due to a factor $\approx$50 increase of peak ground acceleration estimates over most of the asteroid surface compared to \cite{Richardson2005} study, impact events triggering seismic activity are, therefore, at least 10$^2$ times more frequent than assumed during previous studies. 
Despite the large error bars on the impact rate of objects more massive than 10 grams, kilometric scale asteroids are expected to experience several of these impacts during one orbit.

\subsection{Source scaling}
The peak ground accelerations and velocities obtained in this study scale linearly with seismic moment.
The seismic moment is almost proportional to the impactor kinetic energy \citep{Walker2004b}.
However, the peak accelerations also depend on the source duration because short sources of small impactors concentrate the energy in a short duration, and generate high accelerations at high frequencies.
Scaling relations presented by \cite{Lognonne2009} suggest that the source duration scales as $m^{\frac{1}{3}}$ with $m$ being the impactor mass.
So, the results presented here can be extrapolated with confidence to within a factor of 10 of the asteroid mass.
\newline
In the case of explosive sources, the seismic moment is fairly well known, but source durations are much shorter than for impacts \citep{Walker2004b}.
So, the extrapolation of the curves presented in Figure \ref{simu_sismo2} to explosive sources is not possible.

\subsection{Limitations and uncertainties}
Our estimates of peak ground accelerations were constructed in order to be a lower bound.
However, in addition to uncertainties related to the quantification of the seismic source, strong variations are expected due to the internal structure model of the asteroids.
The amount of seismic scattering is a key parameter, because if the seismic scattering is too high the direct waves quantified in this study may not generate the highest peak accelerations.
The shape of the asteroid will also strongly focus/defocus waves, and major fractures may block the waves \citep{Walker2006,Martin2008}.
These two effects create strong variations of peak ground accelerations on the surface.
Eventually, regions with low cohesion may experience irreversible deformations induced by seismic waves, breaking the continuum mechanics hypothesis implicit to our computations.
This effect is not significant at low impactor energy, but is predominant at high impactor energy \citep{asphaug08}.

\begin{figure}
(a) \includegraphics[width=0.51\linewidth,draft=false]{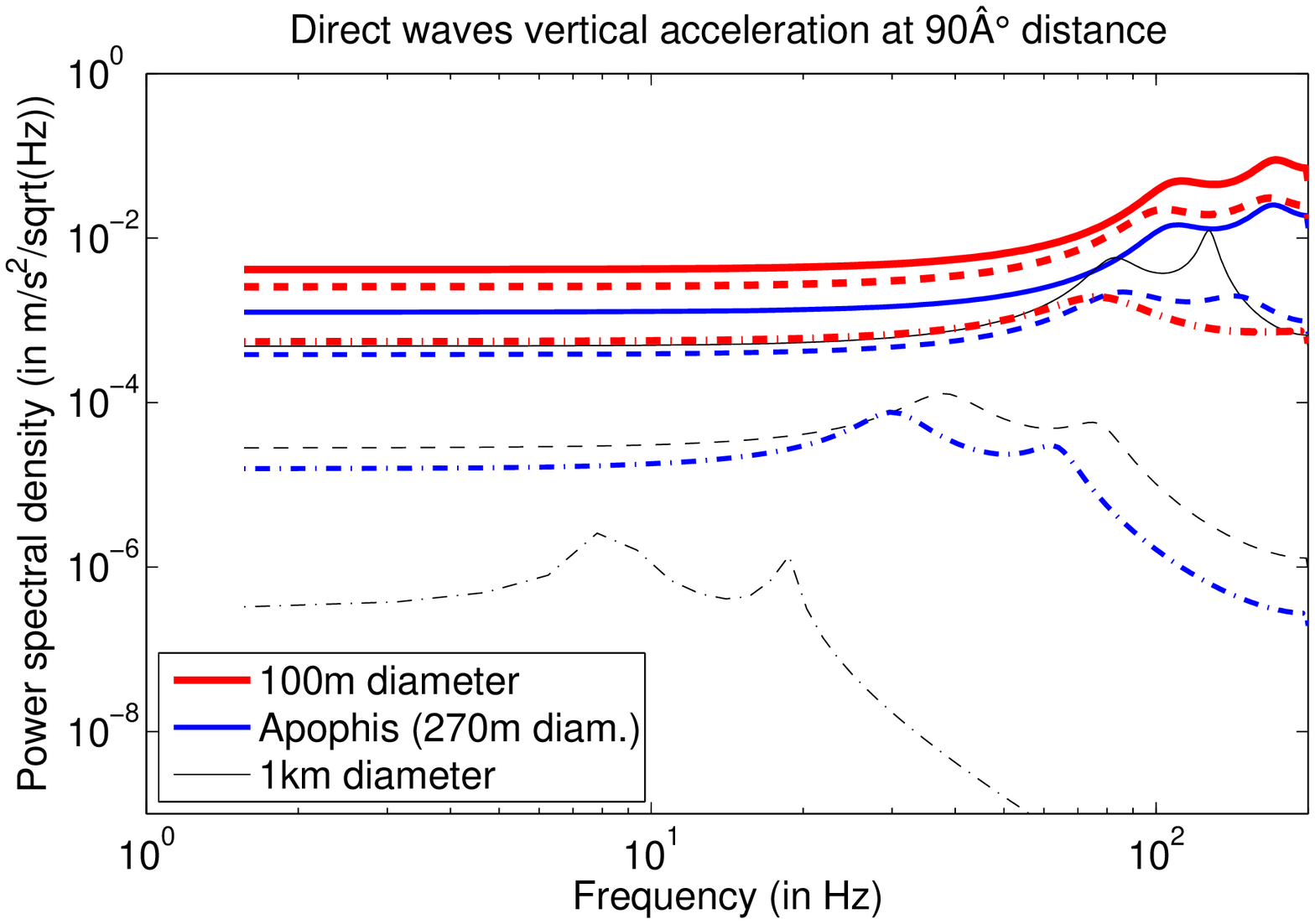}
(b) \includegraphics[width=0.39\linewidth,draft=false]{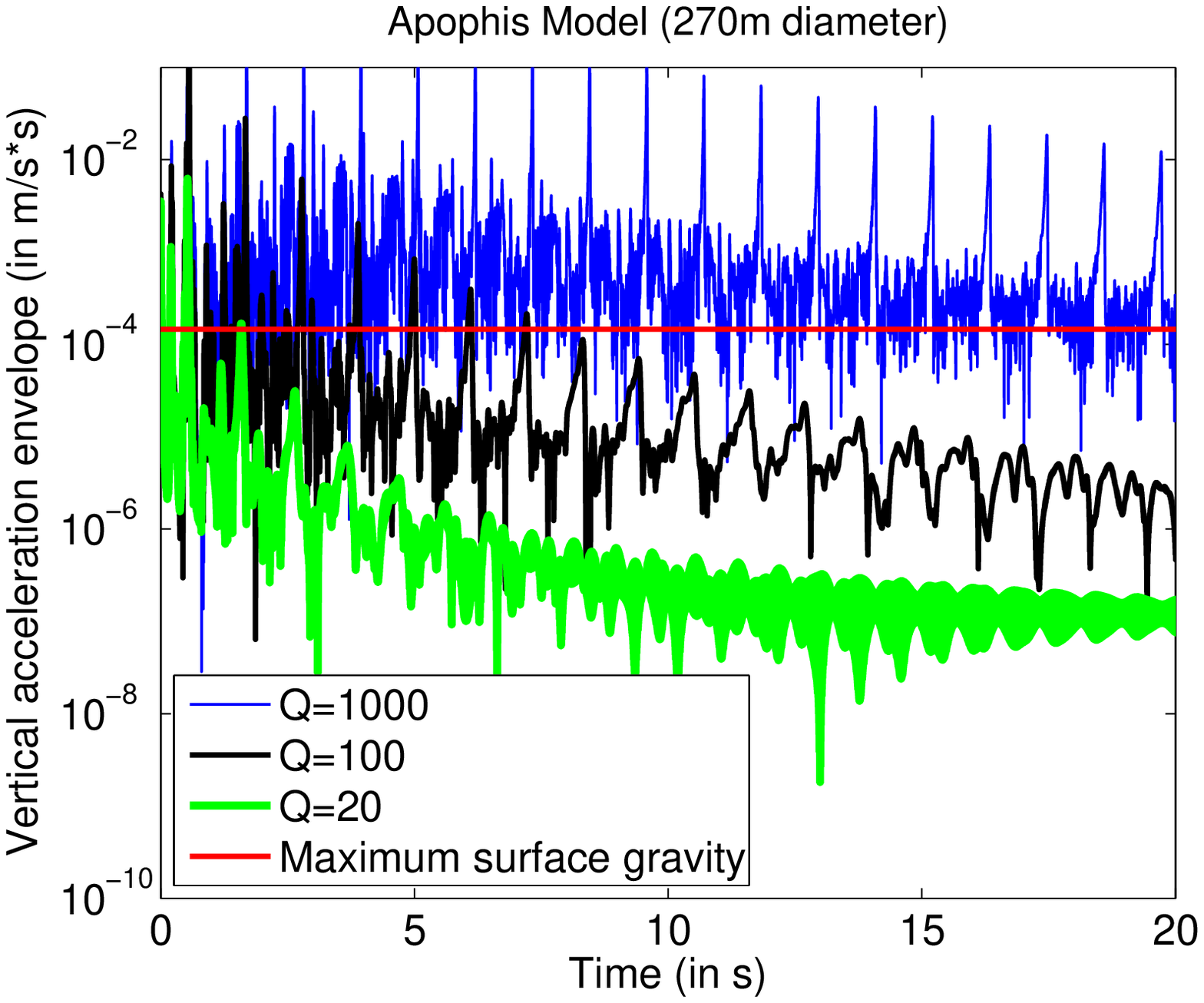}
\caption{ On the left, power spectral densities of acceleration impulse responses of direct body and surface waves at 90$^{\circ}$ epicentral distance to an impact of a 10g object at 6 km/s for homogeneous asteroid models of varying diameters of 100 m (red thick lines), 270 m (blue lines) and 1 km (black thin line).
Computations are performed through normal mode summations as described below for different values of the quality factor of 1000 (plain lines), 100 (dashed lines) and 20 (dotted lines).
On the right, envelope of surface accelerations for an object of 270 m diameter for different quality factors (1000, 100, 20).
}
\label{PSD}
\end{figure}

\begin{figure}
(a)\includegraphics[width=0.9\linewidth,draft=false]{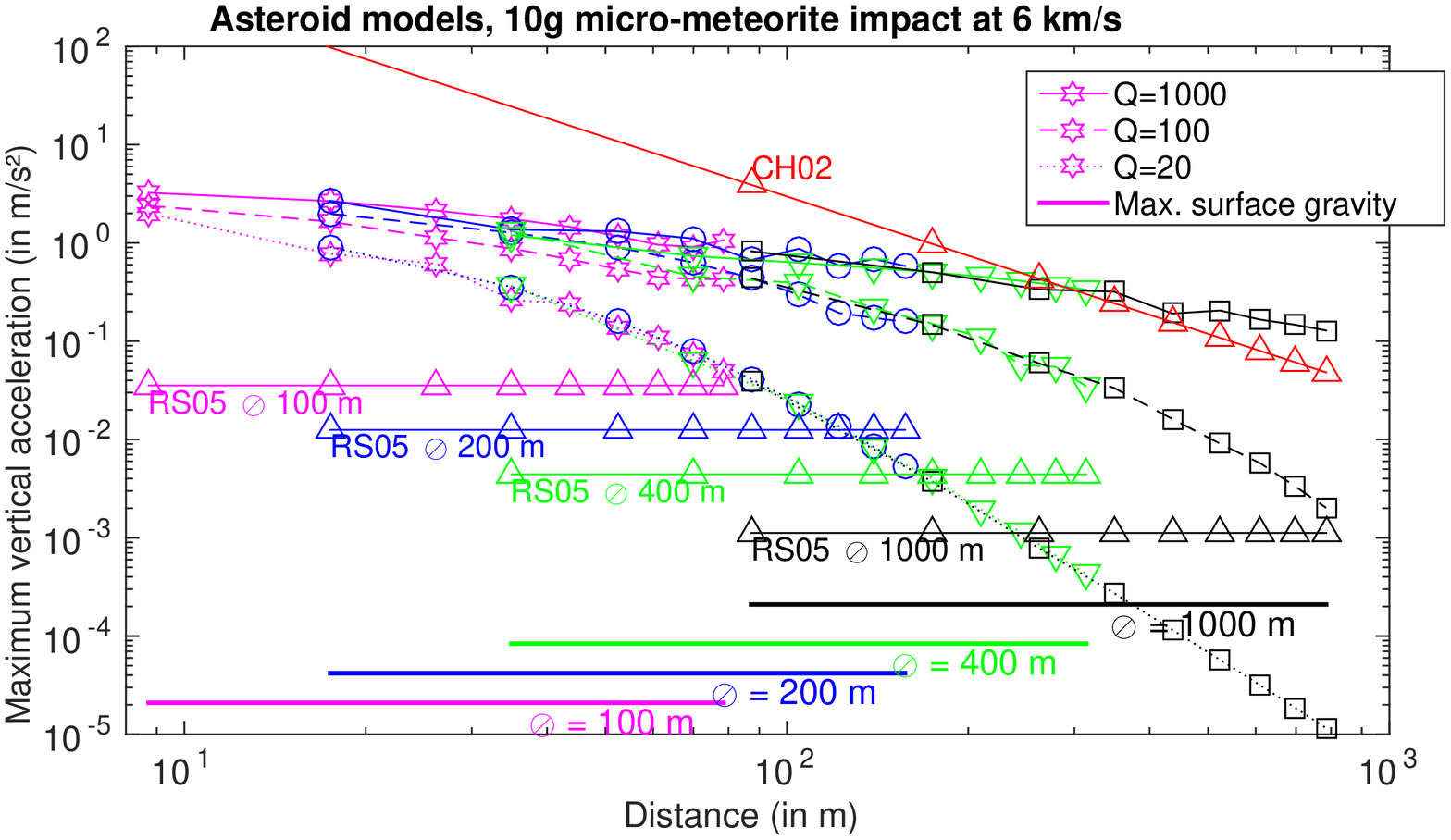}
\newline
(b)\includegraphics[width=0.9\linewidth,draft=false]{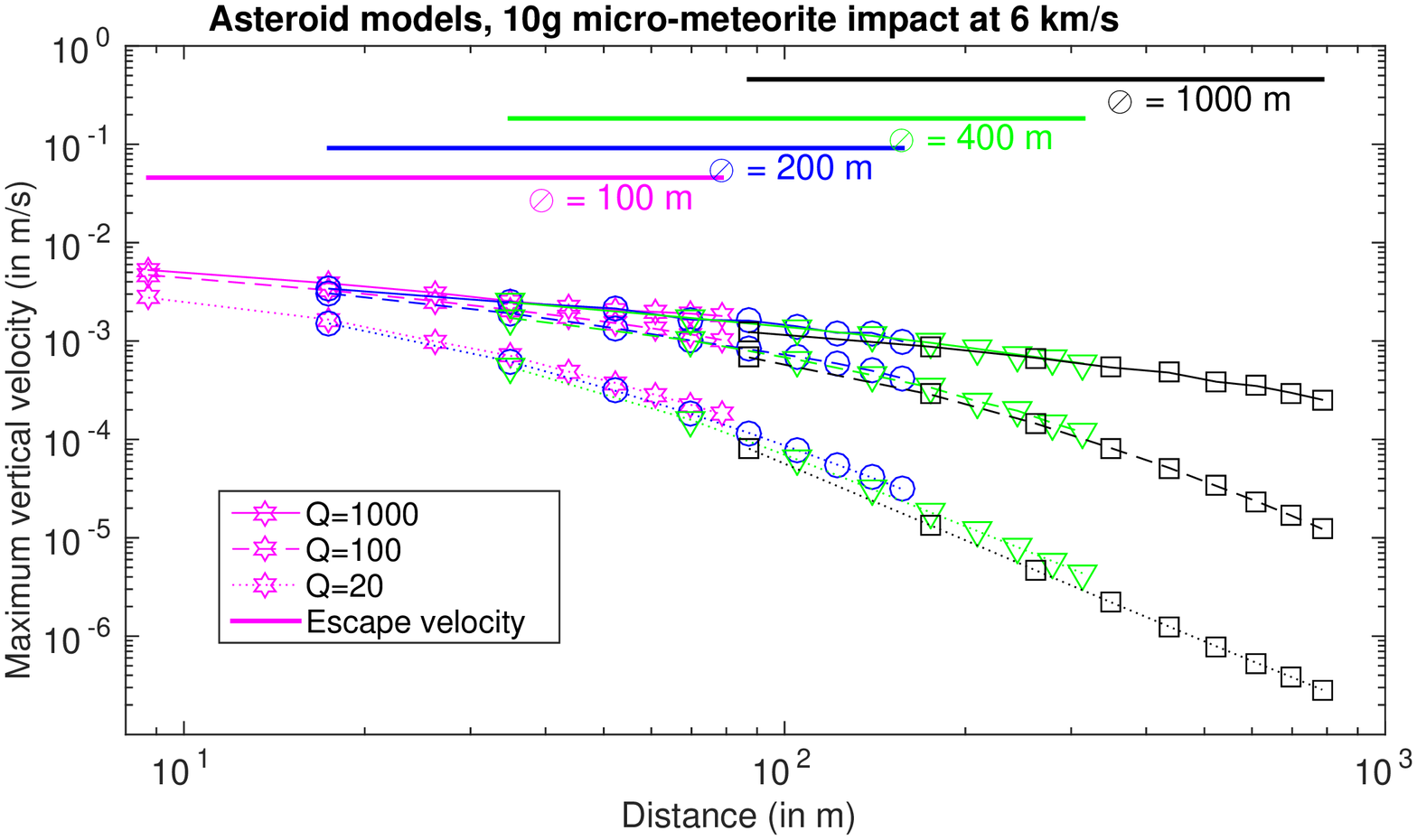}
\caption{Simulations of peak ground accelerations (a) and peak ground velocities (b) as a function of distance (in m) to an impact of a 10g object at 6 km/s for homogeneous asteroid models of following diameters: 100 m (magenta lines), 200 m (blue lines), 400 m (green lines) and 1 km (black lines); and following quality factors: 1000 (plain lines), 100 (dashed lines) and 20 (dotted lines).
In panel (a), the maximum surface gravity of the various asteroid models (thick lines labelled by asteroid diameters), and previous estimates presented by \cite{Cheng02} (CH02) and \cite{Richardson2005} (RS05) are also presented.
In panel (b), the escape velocity of the various asteroid models (thick lines labelled by asteroid diameters) are also presented, assuming that the asteroids are non-rotating.}
\label{simu_sismo2}
\end{figure}


\section{Regolith mobility}

The resulting motion of the non-cohesive regolith layer due to the seismic activity may involve different types of behaviour. The first of these is a ballistic behaviour when the regolith particles are temporarily lofted from the surface. Lofting occurs when the vertical acceleration of the ground is larger than the ambient gravity  \citep[the gravitational attraction of the asteroid on a grain and the inertial effects that arise due to the rotation of the asteroid; ][]{Scheeres2010}.  However, the regolith layer will only be truly lofted if its initial vertical velocity at the instant of ejection (given by the vertical velocity of the ground at the instant of ejection) is sufficient for the particle to have a larger vertical displacement than the ground. Once lofted, we assume that the regolith layer follows a ballistic trajectory until returning into contact with the ground unless, of course, the initial vertical velocity is larger than the escape velocity in which case the regolith will be lost (note that in these simulations this never occurs; see Fig. \ref{simu_sismo2} [b]). The regolith layer may be lofted several times for a given seismic event and, due to the low-gravity environment, even when the particles are lofted to very small heights, the period of free fall can be long. 

\begin{figure}
\includegraphics[width=0.9\linewidth,draft=false]{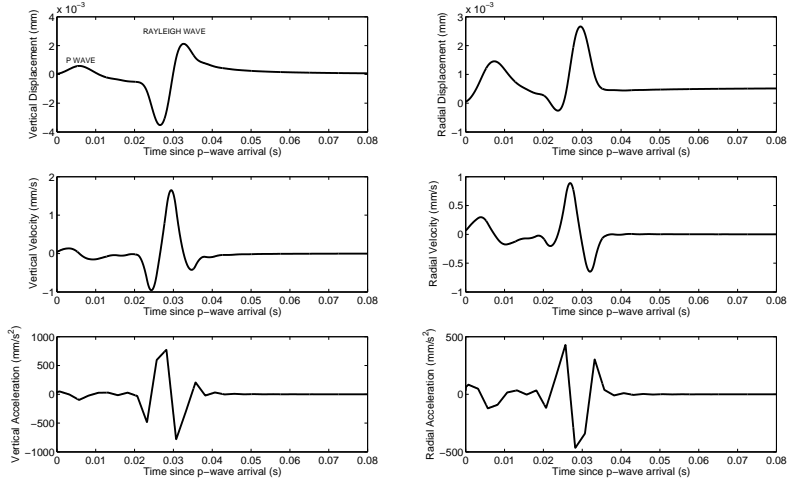}
\caption{Vertical (on the left) and horizontal (on the right) seismic signals computed by our simulations. The example shown here is the seismic signal measured at a distance of 17.5 m away from the micro-meteoroid impact on the surface of an asteroid of diameter 100 m, where Q is assumed to be equal to 20.  The records start at the arrival of the p-wave (the body wave), followed by the arrival of the first Rayleigh (surface) wave approximately 20 ms later.}
\label{signal_ex}
\end{figure}

\begin{figure}
(a)\includegraphics[width=0.9\linewidth,draft=false]{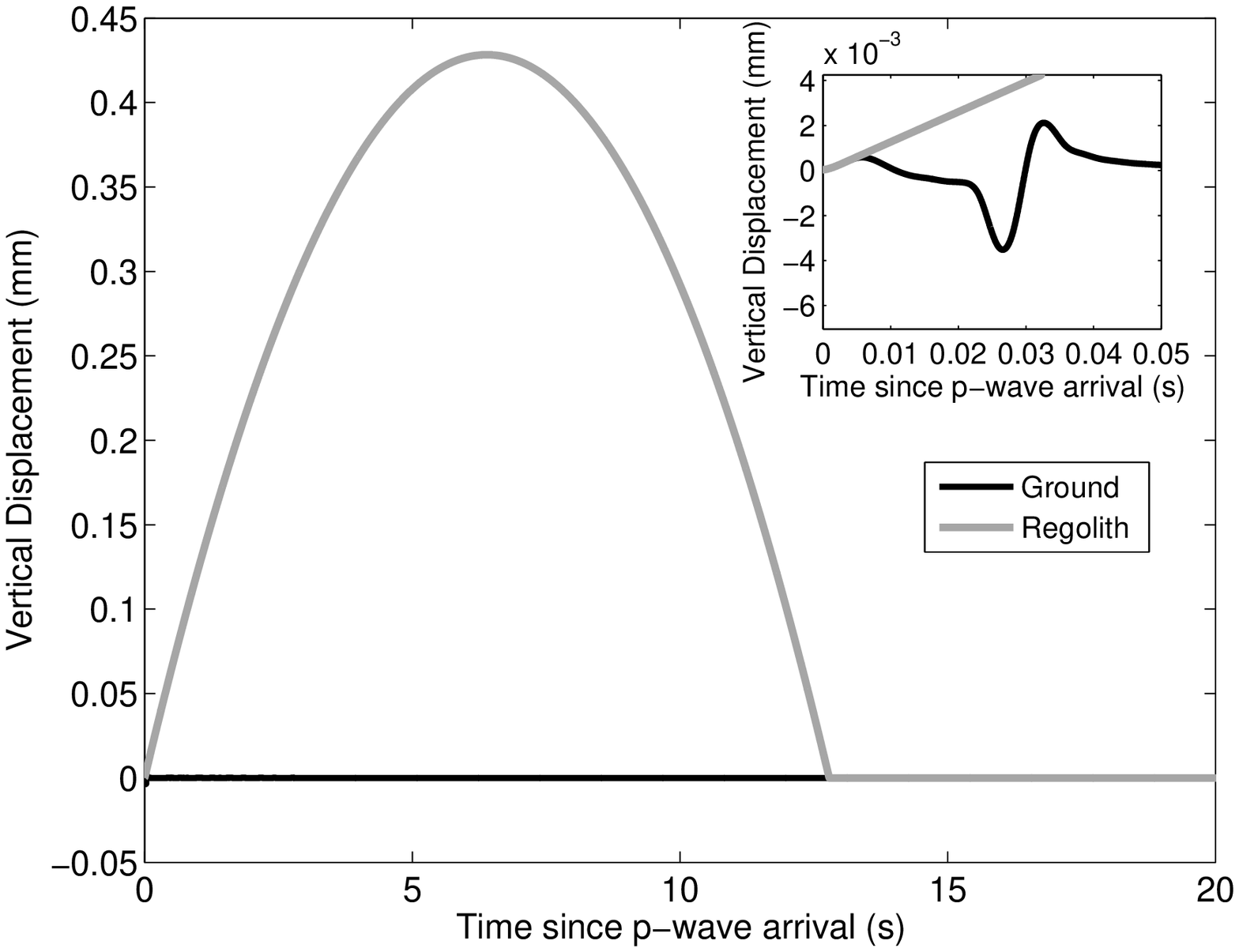}
\newline
(b)\includegraphics[width=0.9\linewidth,draft=false]{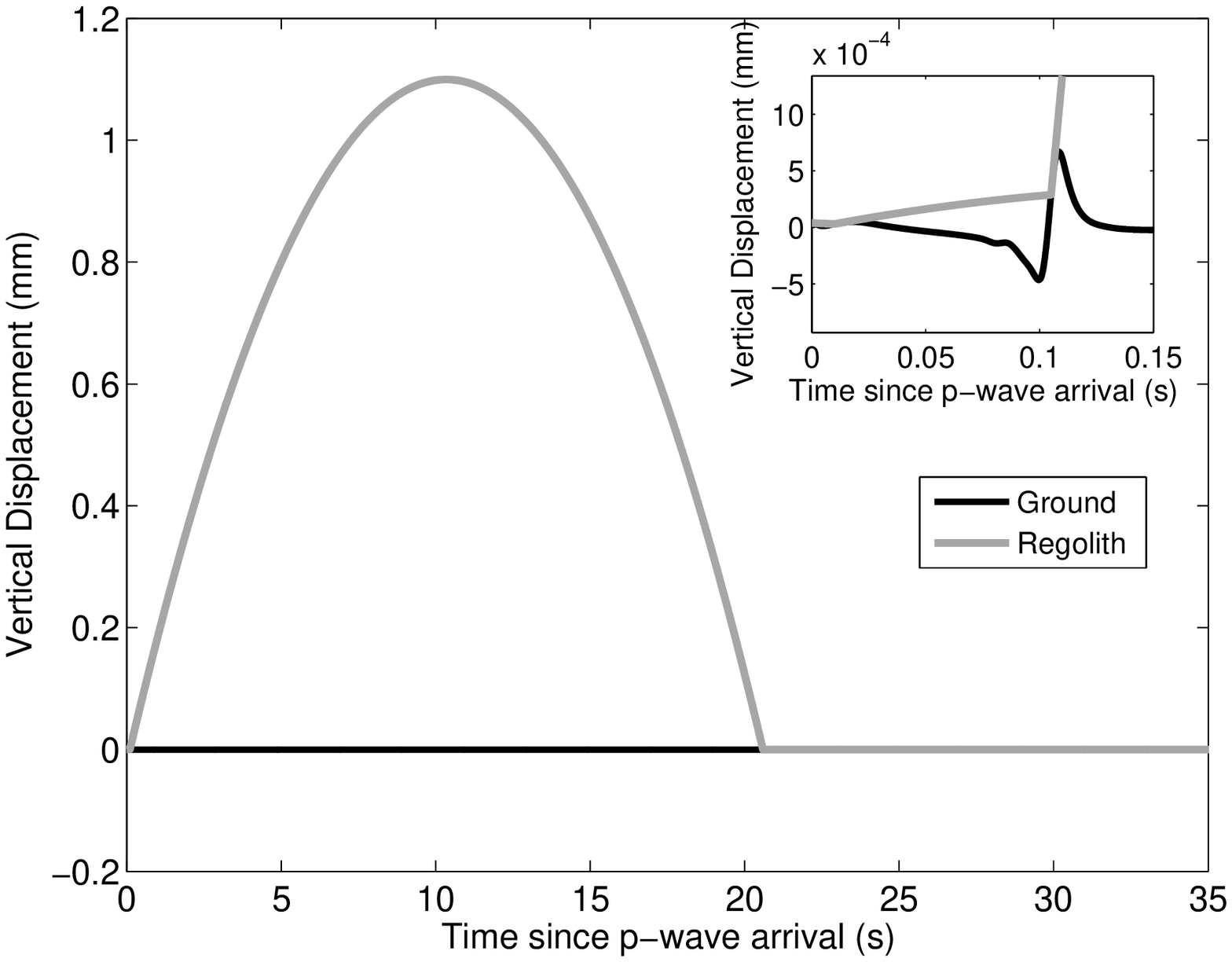}
\caption{Examples of lofting behaviour. (a) The regolith layer is lofted by the p-wave and does not return into contact with the ground until after the Rayleigh wave has passed. (b) The regolith layer is lofted by the p-wave and returns into contact with the ground as the Rayleigh wave is passing and is subsequently relofted.}
\label{lofting_ex}
\end{figure}

Shown in figure \ref{signal_ex} is the resulting ground displacement, velocity and acceleration in both the vertical and the horizontal (radial) directions for an example seismic simulation.  The arrival of the p-wave (body wave) and the Rayleigh (surface) wave can be seen in the figure. For the range of seismic simulations presented above we find three different cases for the ballistic regolith motion: (1) the regolith layer is lofted by the p-wave and does not return into contact with the ground until after the Rayleigh wave has passed (Fig. \ref{lofting_ex} [a]), (2) the regolith layer is lofted by the p-wave and returns into contact with the ground before or as the Rayleigh wave is passing and is subsequently re-lofted (Fig. \ref{lofting_ex} [b]), and (3) the regolith layer is never lofted. 

\begin{figure}
\includegraphics[width=0.9\linewidth,draft=false]{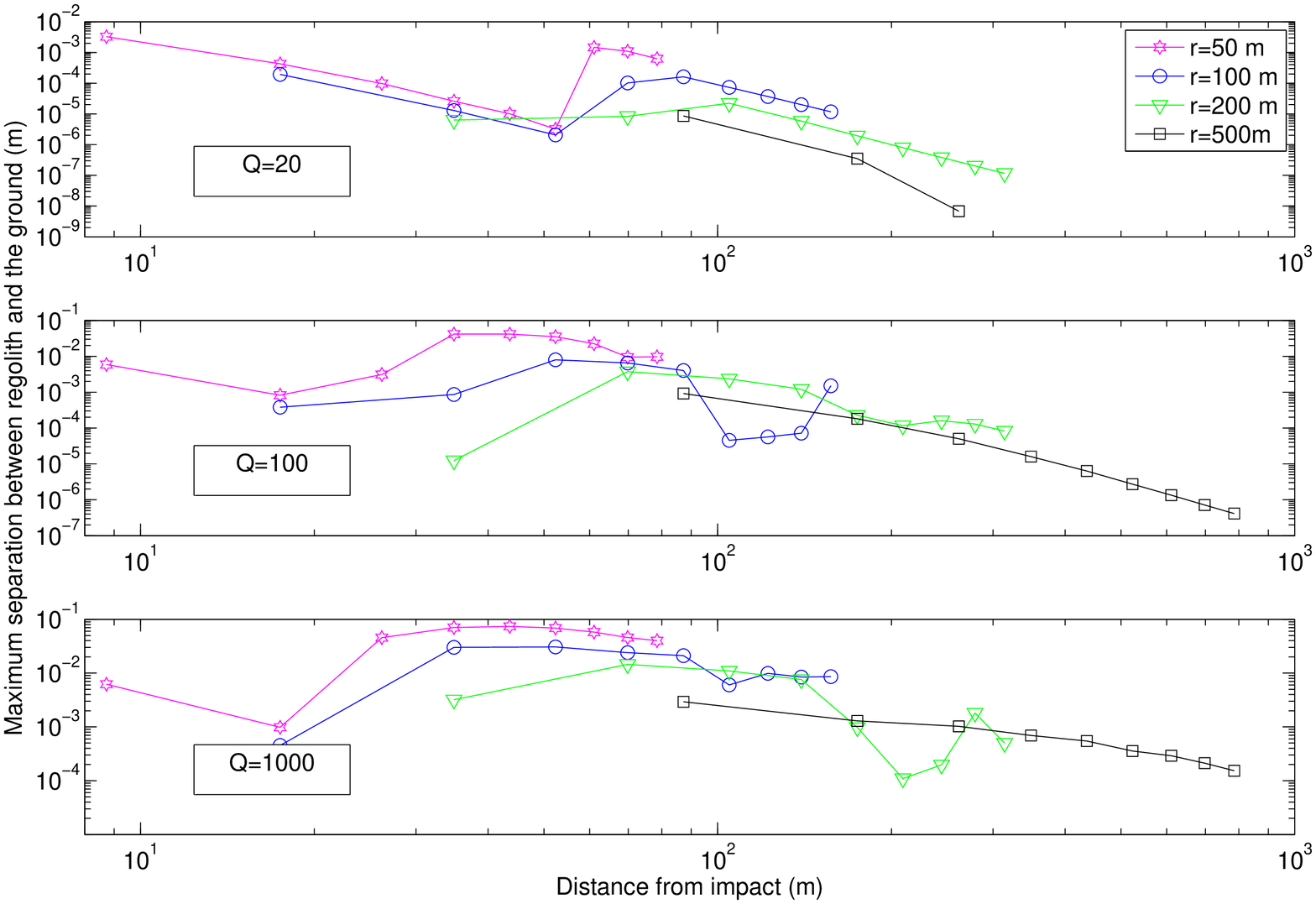}
\caption{Separation between ground and regolith during lofting - The maximum separation achieved between the regolith layer and the ground as a result of lofting, as a function of distance (in m) to an impact of a 10g object at 6 km/s for homogeneous asteroid models of following diameters: 100 m (magenta lines), 200 m (blue lines), 400 m (green lines) and 1 km (black lines); and following quality factors: 20 (top), 100 (middle) and 1000 (bottom). The slope with respect to the local surface gravity is 0$^{\circ}$ and the asteroids are assumed to be non-rotating.}
\label{separation}
\end{figure}

\begin{figure}
\includegraphics[width=0.9\linewidth,draft=false]{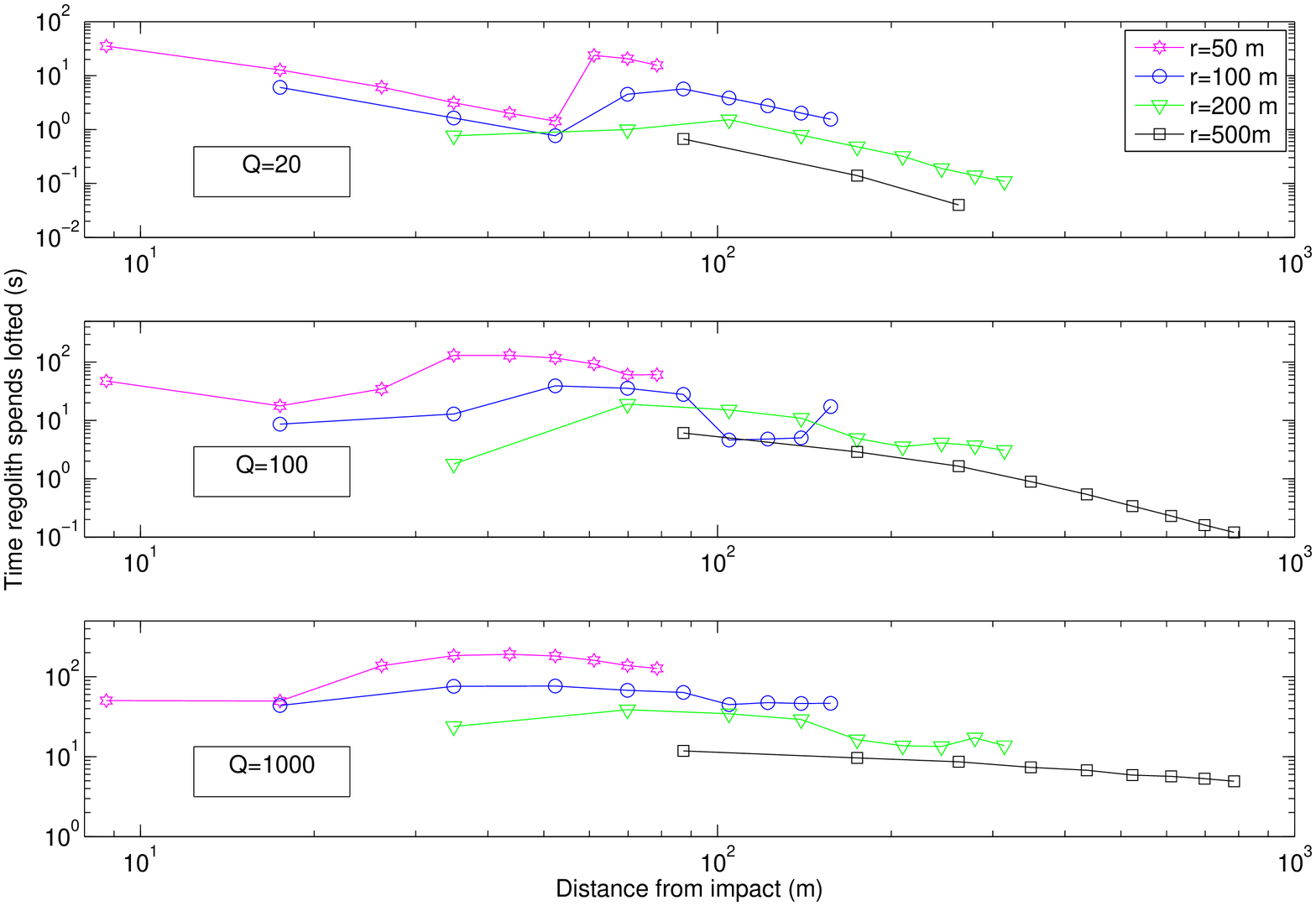}
\caption{Total lofting time - The total time the regolith spends lofted, as a function of distance (in m) to an impact of a 10g object at 6 km/s for homogeneous asteroid models of following diameters: 100 m (magenta lines), 200 m (blue lines), 400 m (green lines) and 1 km (black lines); and following quality factors: 20 (top), 100 (middle) and 1000 (bottom). The slope with respect to the local surface gravity is 0$^{\circ}$ and the asteroids are assumed to be non-rotating.}
\label{loftingtime}
\end{figure}

As the Rayleigh wave has a much higher amplitude than the p-wave, the particles lofted by this wave will reach larger heights and will spend much more time in free fall than those lofted by the p-wave. This leads to a counterintuitive result that, close to the micro-meteoroid impact, the regolith layer may be lofted less than further from the micro-meteoroid impact. This can be seen in figures \ref{separation} and \ref{loftingtime}, where the maximum lofting height reached, and the total time spent lofted, is shown for the full range of seismic simulations assuming that the slope with respect to the local surface gravity is 0$^{\circ}$ and the asteroids are non-rotating.  Close to the micro-meteoroid impact the regolith is lofted on the p-wave and the lofting time and height decrease as a function of distance from the micro-meteoroid impact as the amplitude of the p-wave decreases.  Then, at a certain distance, the lofting height and time suddenly increase; this corresponds to the distance at which the regolith starts to be lofted on the Rayleigh wave.  This dip is particularly clear for the simulations with high attenuation (low Q). 

\begin{figure}
\includegraphics[width=0.9\linewidth,draft=false]{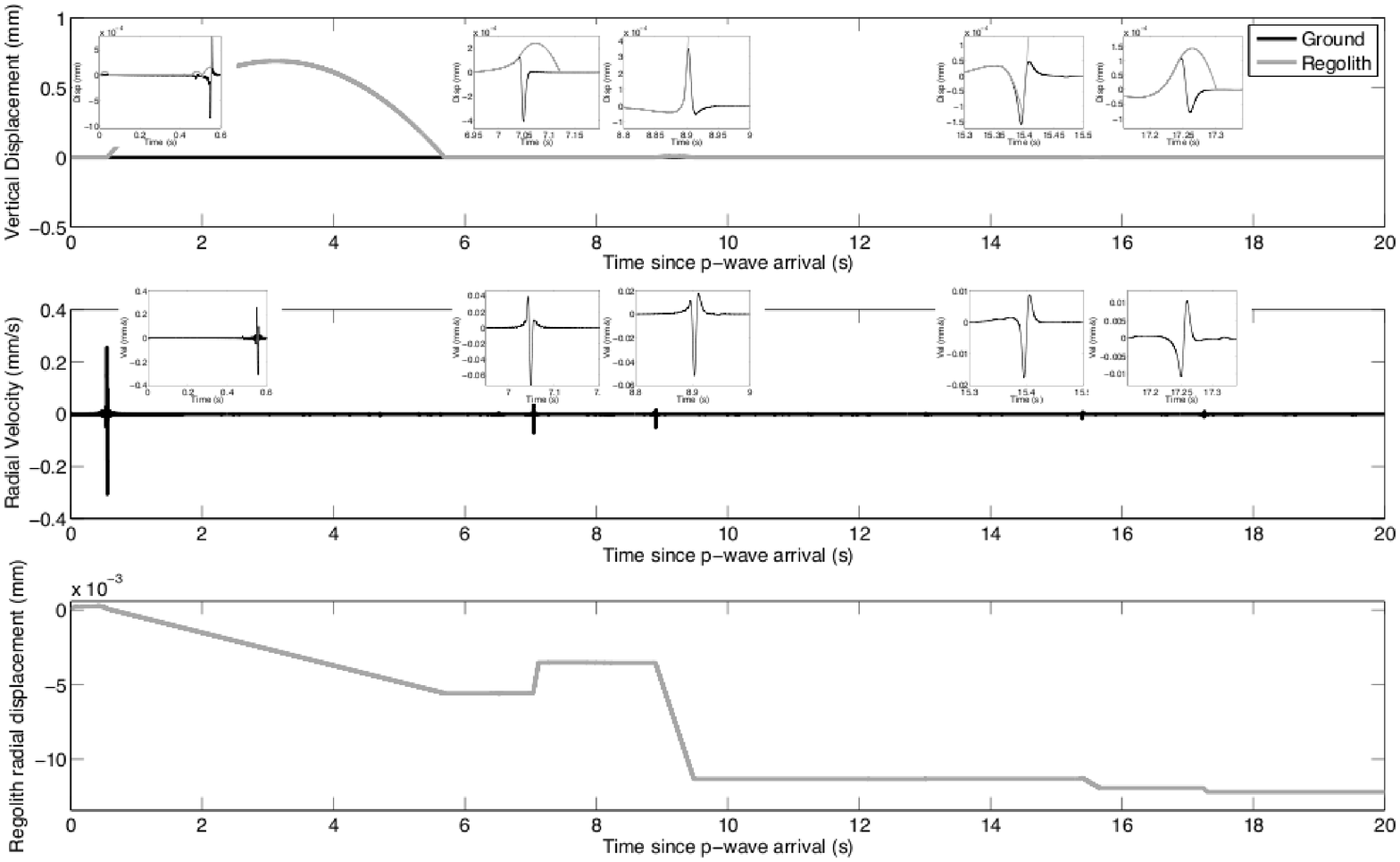}
\caption{Horizontal regolith motion during the ballistic phases - Top: the ground and regolith vertical displacements as a function of time showing the the ballistic periods. Middle: the radial velocity of the ground as a function of time. Bottom: The resulting horizontal displacement of the regolith during the ballistic periods.  Note that the horizontal motion is not always in the same direction as it depends on the sign of the ground radial velocity at the moment of lofting. In this example the asteroid is 1 km in diameter, Q = 1000, the epicentral distance from the micro-meteoroid impact is 40$^{\circ}$ (700 m distance), the slope with respect to the local surface gravity is 0$^{\circ}$ and the asteroid is non-rotating.}
\label{horizontal_motion}
\end{figure}

Note that the seismic simulations last 50 seconds and, for the lofting calculations, we take the assumption that at times greater than 50 seconds the ground is stationary (this is a worst case estimate, especially for the low attenuation simulations).  The ballistic regime is generally dominated by one long period of lofting (typically several 10s of seconds) that can be followed (and sometimes preceded) by several shorter jumps (see Fig. \ref{horizontal_motion}).

\begin{figure}
\includegraphics[width=0.9\linewidth,draft=false]{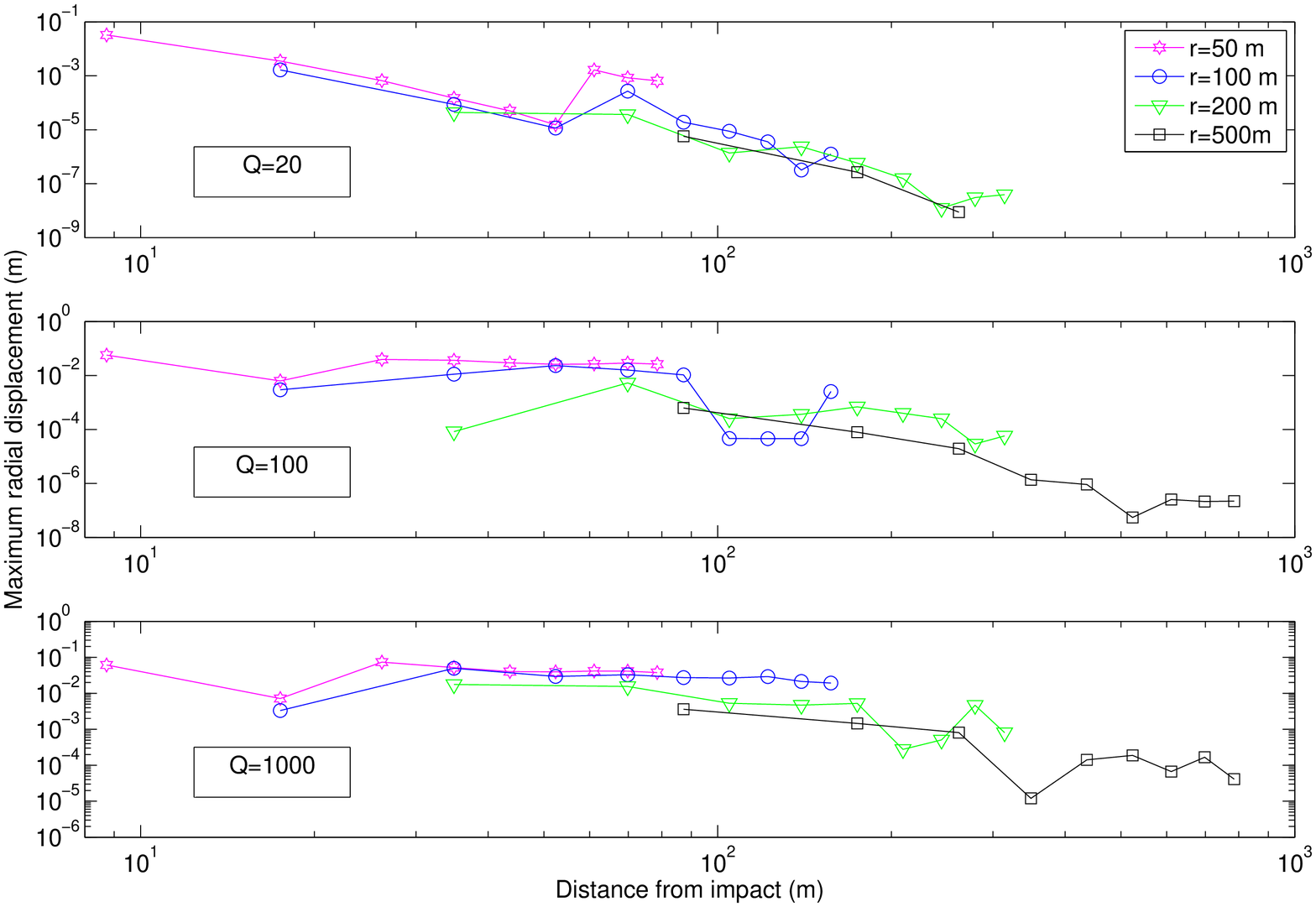}
\caption{Regolith radial displacement -  The total radial displacement of the regolith during lofting , as a function of distance (in m) to an impact of a 10g object at 6 km/s for homogeneous asteroid models of following diameters: 100m (magenta lines), 200 m (blue lines), 400m (green lines) and 1 km (black lines); and following quality factors: 20 (top), 100 (middle) and 1000 (bottom). It is assumed that the slope with respect to the local surface gravity is 0$^{\circ}$ and the asteroid is non-rotating.}
\label{displacement}
\end{figure}

 If the slope with respect to the local surface gravity is 0$^{\circ}$, the horizontal motion of the regolith layer during the ballistic trajectory can also be estimated assuming that the regolith has a constant horizontal velocity determined by the horizontal (radial) velocity of the ground at the instant of ejection.  However, the horizontal motion will not always be in the same direction (see example in Fig. \ref{horizontal_motion}). The total horizontal displacement for the regolith layer during the ballistic phase for all of the seismic simulations is given in figure \ref{displacement}, assuming that the slope with respect to the local surface gravity is 0$^{\circ}$ and the asteroids are non-rotating. For the low attenuation (Q = 1000) seismic simulations, the maximum horizontal displacements due to ballistic motion vary from $\sim$6 cm at distances $<$10 m from the micro-meteoroid impact to $\sim$0.01 mm at distances $>$300 m from the micro-meteoroid impact. 

\begin{figure}
\includegraphics[width=0.9\linewidth,draft=false]{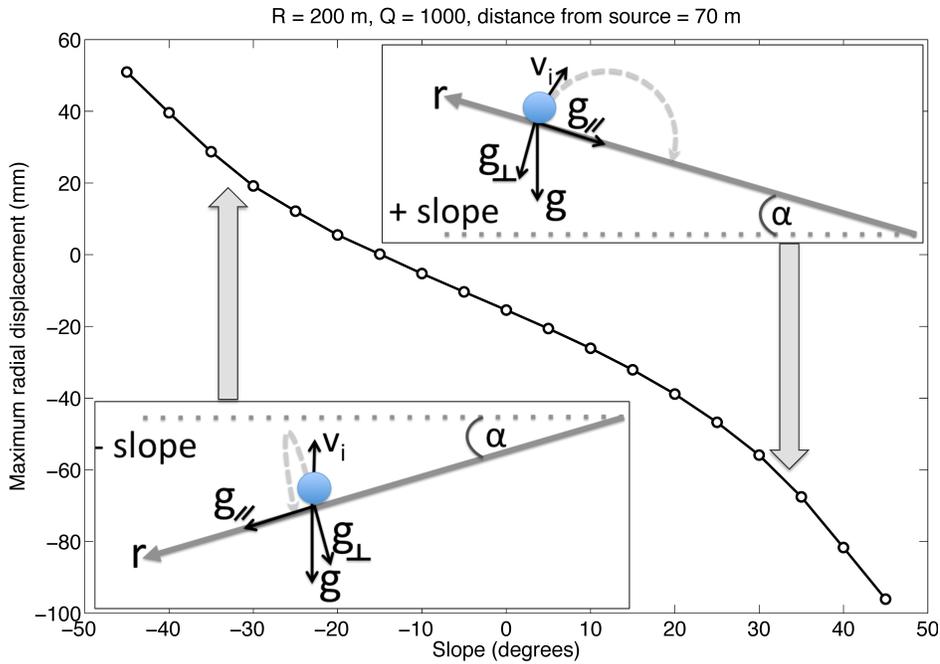}
\caption{Influence of slope on regolith radial displacement -  The total radial displacement of the regolith during lofting as a function of the slope with respect to the local surface gravity (\bf{$g$}). The example given is for regolith at a distance of 70 m from a micro-meteoroid impact on the surface of a non-rotating asteroid of diameter 400 m. The quality factor is 1000. The schematics explain the motion of the regolith during lofting. The initial velocity (\bf{$v_i$}) is identical in all cases (with respect to the slope), and the subsequent trajectory is strongly influenced by the  direction of the local surface acceleration, favouring downslope motion.}
\label{downslope_motion}
\end{figure}

If, however, the slope with respect to the local surface gravity is non-zero, the motion of the regolith layer during the ballistic trajectory will be strongly influenced by the direction of the local surface acceleration. As the component of the local surface gravity perpendicular to the slope will be reduced, the lofting times will be longer. Additionally, the component of the local surface gravity parallel to the slope will favour downslope horizontal motion (see Fig. \ref{downslope_motion}).

Our lofting calculations have shown that, during even a small micro-meteoroid impact (10g at 6 km/s) the non-cohesive regolith layer on small (sub-kilometric) asteroids is uplifted but not lost.  The initial motion of the regolith layer will be dictated by the ground velocity at the moment of ejection.  However, the following ballistic motion is strongly influenced by the direction of the ambient surface acceleration relative to the surface, favouring downslope regolith displacements.

Similarly, there are other types of behaviour that may contribute to the radial motion.  When the regolith layer is in contact with the ground and is subjected to horizontal motion coming from the ground (i.e., applied at the base of the regolith layer) downslope motion will occur in a stop-start fashion, referred to as the {\it stick-slip} motion by \cite{Richardson2005}.  However, to understand the regolith motion the simple sliding block analysis \citep{newmark65, Richardson2005,jibson00} is not sufficient because the seismic wavelengths in our simulations are likely to be of similar size, or perhaps even smaller than the depth of the regolith layer (see Fig. \ref{PSD}; $V_P = 900 m/s$).  Acoustic fluidisation of granular materials by the seismic waves \citep{melosh1979} may also increase regolith motion in the same, downslope, direction. More detailed numerical simulations are required to study these types of regolith motion and any regolith flow that may occur as the regolith layer comes into contact with the ground at the end of a ballistic period.  However, it is clear that over time, the cumulative effect of these mechanisms will result in accumulations of regolith material in surface acceleration potential lows or, ``potential valleys'', as has been observed on several asteroids. On the surface of Itokawa, for example, the Muses Sea and Sagamihara regions coincide with the potential valleys and are generally homogeneous, and relatively flat (slopes $<8^\circ$), consistent with a regolith layer that has been allowed to seek out its minimum energy configuration after the formation of the asteroid \citep{Miyamoto2007, fujiwara2006,yano06}.

\section{Implications of regolith concentration}

\begin{figure}
(a) \includegraphics[width=0.45\linewidth,draft=false]{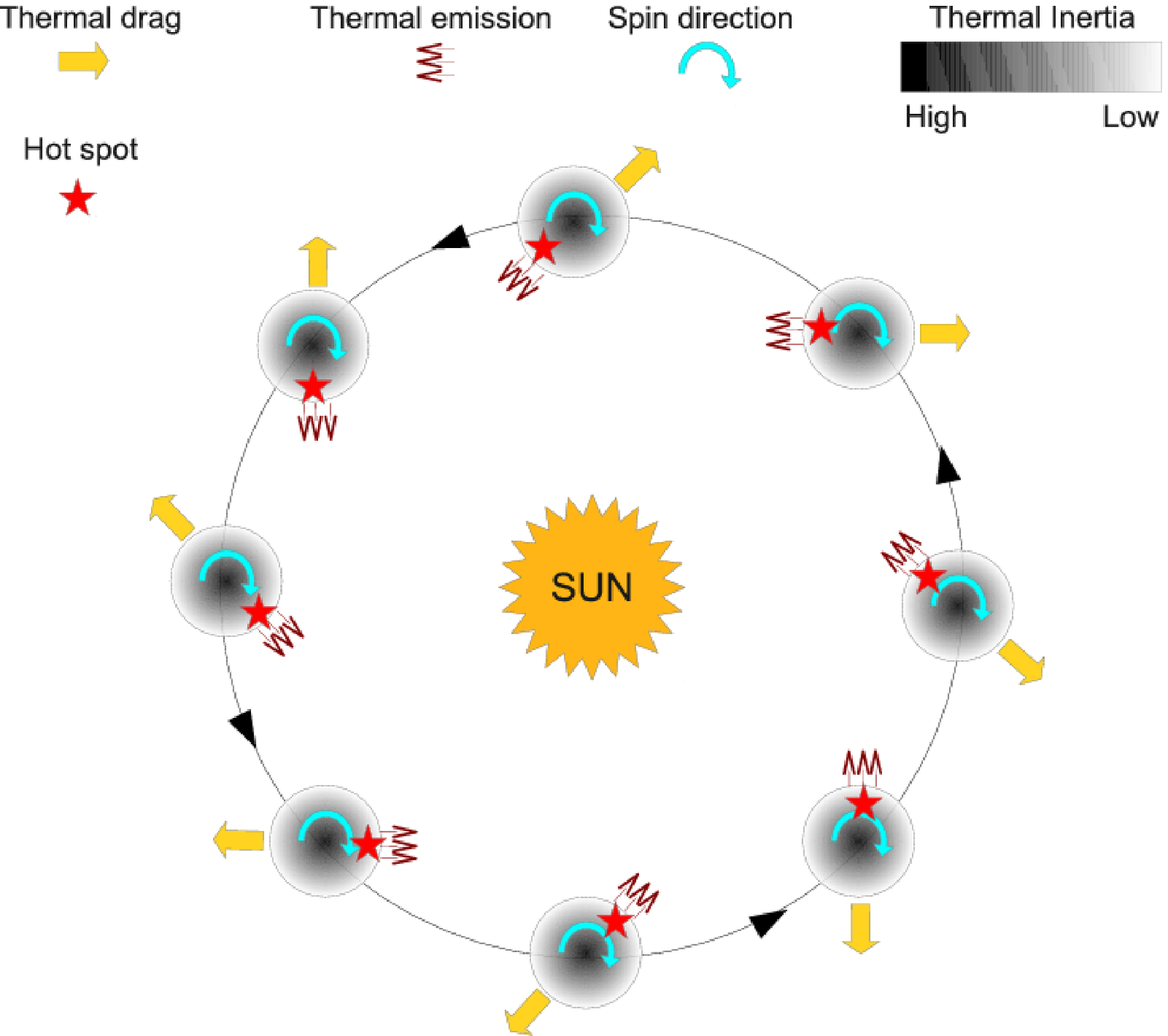}
(b) \includegraphics[width=0.45\linewidth,draft=false]{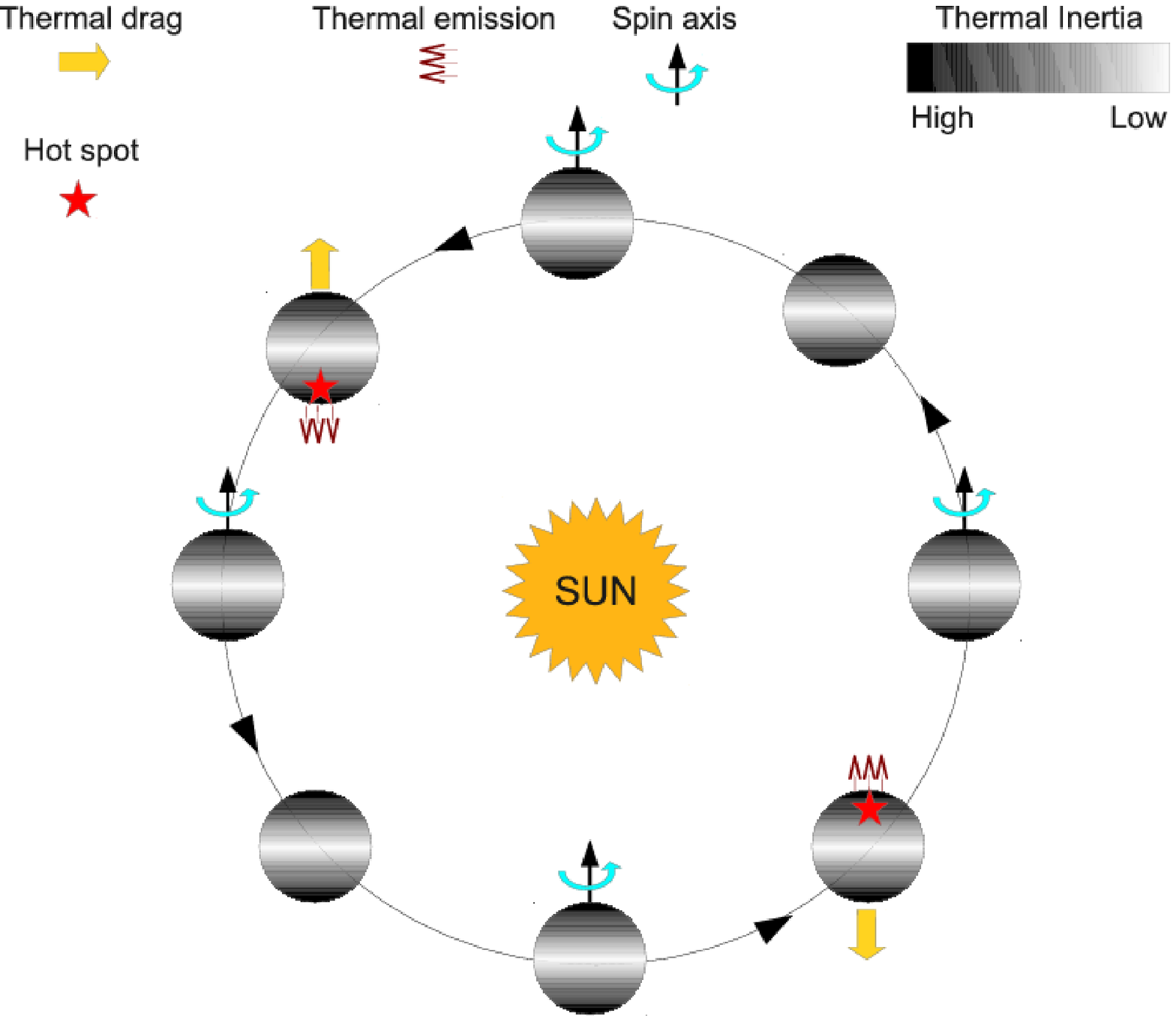}
\caption{Sketches of diurnal (a) and seasonal (b) Yarkovsky effects with latitudinal dependence of thermal inertia. (a) obliquity angle is 180$^{\circ}$ (retrograde). (b)  obliquity angle is 90$^{\circ}$. Adapted from \cite{Bottke2006}.}
\label{Yarko}
\end{figure}

The non-cohesive regolith layer has a thermal inertia 3 to 4 orders of magnitude smaller than bare rocks forming the cohesive part \citep{Delbo2007}. 
Our modelling predicts that, for a given source (here a micro-meteoroid impact), the induced surface accelerations tend to increase as the asteroid size decreases due to a decrease of the average distance to the source. Smaller asteroids will also have longer lofting times resulting in an increased mobility of the loose regolith towards regions of low surface acceleration potential. However, as already pointed out above, nonlinearities in both the asteroid response to impacts and the uplift phenomena prevent a simple scaling with asteroid size.
\newline
The high mobility of uncohesive regolith may expose an increasing number of bare rocks of high thermal inertia thus helping to explain the observed trend of increasing average thermal inertia with decreasing asteroid size \citep{Delbo2007}; because the thermal inertia ratio between fine dust and bare rocks is so large, the surface average is strongly influenced by high thermal inertia regions.
\newline
As a companion effect, the concentration of surface regolith in specific areas would create large lateral variations of thermal inertia from pole to equator that may influence the Yarkovsky effect on these objects \citep{Rubincam1995,Farinella1998,Bottke2006}.
In the particular case of small fast rotators, the regolith will concentrate in equatorial areas (surface acceleration potential lows), and consequently give this region a low surface thermal inertia. The sensitivity of such asteroids to the diurnal Yarkovsky effect may, therefore, be significant because the thermal energy will be re-emitted by the surface in a short delay comparable to the rotation period (Fig. \ref{Yarko}(a)). Moreover, the polar areas will be depleted in regolith and, consequently, will have a larger surface thermal inertia. Therefore, these objects may also be sensitive to the seasonal Yarkovsky effect, because the time necessary to re-emit the thermal energy is long, and comparable to the orbit period (Fig. \ref{Yarko}(b)).
Consequently, even if the Yarkovsky effect depends on the pole orientation of the asteroid, {\bf smaller and faster} rotators may have an increased orbital mobility due to surface thermal inertia variations induced by fine grained regolith concentration in equatorial areas.
{\bf
However, the significance of this effect as a function of asteroid size and rotation speed should be studied in more detail.
}
\newline
In addition, an increase in regolith migration following micro-meteoroid impacts may enhance the efficiency of regolith production via both impacts and thermal fatigue \citep{Delbo2014} by regularly relocating the freshly produced regolith to the potential lows and, in turn, exposing more material to the surface thermal environment and to impacting meteoroids.
\newline
The results presented here also have implications for lander missions on the surface of these small bodies, because a rough estimate of the regolith thickness is important for landing or penetration dynamics. In addition, as the lofting time computations presented in figure \ref{loftingtime} do not depend on the lofted mass, these computations can be applied to any lander poorly coupled to the asteroid surface. Therefore, the ground coupling capabilities and active source types of future seismic experiments are also affected by our results, which suggest that active sources may uplift uncoupled landers during a duration long enough to miss most of the seismic signal.

\section{Conclusion}

Assuming that asteroid interiors are cohesive enough to support the propagation of the seismic wave generated by micro-meteoroid impacts without irreversible deformations, continuum mechanics simulations of seismic wave propagation are presented for asteroids in the kilometric size range. Despite conservative hypotheses on the source amplitude and wave attenuation, we find that extensive seismic activity can be activated by the impact of even small micro-meteoroids. Therefore, seismic activity generated by impacts is much more frequent than assumed during previous studies.

Additionally, the high frequency surface accelerations generated by the direct seismic waves exceed the maximum surface gravity at a global scale for micro-meteoroid impactors (10g at 6 km/s). Therefore, during such events, any loose, uncohesive regolith at the surface of the asteroid is uplifted but not lost.  The lofting times and displacements are estimated assuming a simplified ballistic behaviour as a first approximation and they present a non-linear behaviour depending on the seismic wave at the origin of the lofting process.  During lofting the loose regolith is  preferentially transported to regions of low surface acceleration potential. 

The concentration of fine grained regolith in specific areas has important implications for the thermal behaviour of the object. For example, the orbital mobility of small fast rotators may be increased due to an increased sensitivity to both diurnal and seasonal Yarkovsky effects. We also present other implications for the deployment of landers or seismic experiments on the surface of these objects.
\newline
Numerous possible implications of non-cohesive regolith uplift and concentration induced by micro-meteoroid impacts on kilometric-sized asteroids have been discussed. However, this work can be continued in many directions: modelling of seismic wave scattering, numerical and experimental characterisation of loose regolith mechanical properties, including cohesive forces in our model that could play an important role in the regolith behaviour \citep{Hartzell2011,Hartzell2013},  modelling lateral variations of thermal inertia and the consequences on the Yarkovsky effect, developing the models to estimate the duration of the shaking following an impact and performing a detailed study of the timescales involved in global redistribution of regolith (and crater erasure) given these predicted seismic surface accelerations and velocities.

%

\section{Acknowledgments}
We acknowledge J.E. Richardson and an anonymous reviewer for their constructive comments on the paper.

\bibliographystyle{model2-names}
\bibliography{article}








\end{document}